\documentclass[apj]{emulateapj-rtx4}

\usepackage{calc}

\shorttitle{A Blue Source at the Location of SN~2011dh}
\shortauthors{Folatelli et al.}

\begin{document}

\title{A Blue Point Source at the Location of Supernova~2011dh}

\author{%
  Gast\'on Folatelli\altaffilmark{1},
  Melina C.\ Bersten\altaffilmark{1},
  Omar G.\ Benvenuto\altaffilmark{2,3},
  Schuyler D.\ Van Dyk\altaffilmark{4},
  Hanindyo Kuncarayakti\altaffilmark{5,6},
  Keiichi Maeda\altaffilmark{7,1},
  Takaya Nozawa\altaffilmark{8},
  Ken'ichi Nomoto\altaffilmark{1,9},
  Mario Hamuy\altaffilmark{6,5},
  and
  Robert M.\ Quimby\altaffilmark{1}
}
\altaffiltext{1}{Kavli Institute for the Physics and Mathematics of
  the Universe (WPI), The University of Tokyo, Kashiwa, Chiba
  277-8583, Japan; gaston.folatelli@ipmu.jp}
\altaffiltext{2}{Facultad de Ciencias Astron\'omicas y
  Geof\'{\i}sicas, Universidad Nacional de La Plata, Paseo del Bosque
  S/N, B1900FWA La Plata, Argentina}
\altaffiltext{3}{Instituto de Astrof\'isica de La Plata (IALP), CONICET,
  Argentina}
\altaffiltext{4}{Spitzer Science Center/Caltech, Mailcode 220-6,
  Pasadena, CA 91125, USA}
\altaffiltext{5}{Millennium Institute of Astrophysics (MAS), Santiago, Chile}  
\altaffiltext{6}{Departamento de Astronom\'ia, Universidad de Chile,
  Casilla 36-D, Santiago, Chile}  
\altaffiltext{7}{Department of Astronomy, Kyoto University,
  Kitashirakawa-Oiwake-cho, Sakyo-ku, Kyoto 606-8502, Japan} 
\altaffiltext{8}{National Astronomical Observatory of Japan, Mitaka,
  Tokyo 181-8588, Japan} 
\altaffiltext{9}{Hamamatsu Professor}

\setcounter{footnote}{10}

\begin{abstract}
\noindent We present {\em Hubble Space Telescope} ({\em HST})
observations of the field of the Type~IIb supernova (SN) 2011dh in M51
performed at $\approx$1161 rest-frame days after explosion using the
Wide Field Camera 3 and near-UV filters F225W and F336W. A star-like
object is detected in both bands and the photometry indicates it has
negative $(\mathrm{F225W} - \mathrm{F336W})$ color. The observed
  object is compatible with the companion of the now-vanished yellow
  supergiant progenitor predicted in interacting binary models. We
  consider it unlikely that the SN is undergoing strong interaction
  and thus estimate that it makes a small contribution to the observed
  flux. The possibilities of having detected an unresolved light echo
  or an unrelated object are briefly discussed and judged unlikely. 
Adopting a possible range of extinction by dust, 
we constrain parameters of the proposed binary system. In particular,
the efficiency of mass accretion onto the binary companion must be
below 50\%, if no significant extinction is produced by newly formed
dust. Further multiband observations are required in order to confirm the 
identification of the object as the companion star. If confirmed,
  the companion star would already be dominant in the UV--optical
  regime, so it would readily provide a unique opportunity to perform
  a detailed study of its properties. 
\end{abstract}


\section{INTRODUCTION}
\label{sec:intro}

\noindent After decades of sustained progress in stellar
  evolution theory, the connection between different types of
core-collapse supernovae (SNe) and their progenitors remains
partly unknown.
Hydrogen-rich, Type II SNe, have been found to arise from red
supergiant stars, as expected from evolutionary models
\citep{smartt09,vandyk12}. 
Conversely, the origin of
hydrogen-poor SNe (Types Ib, Ic and IIb) is unclear and one
important question is how their progenitors expel the hydrogen
envelope. This process may be regulated by wind, which is
expected to be stronger for larger progenitor mass. However, there is
compelling evidence that binarity and, in particular, mass-transfer
processes in interacting binaries, must play an important role in the
evolution of most massive stars \citep{sana12}. 

In this context, characterizing the progenitors of Type IIb
SNe, which retain a small fraction of their hydrogen envelopes, becomes
particularly relevant. One of the best-studied objects of this type is 
SN~1993J, whose progenitor has been identified as a massive binary
system composed of a K-type supergiant progenitor plus a B-type 
supergiant companion \citep{maund04,maund09,fox14}. More
recently, the discovery of the Type~IIb SN~2011dh in M51 provided the
opportunity---with improved observational capability---of identifying
yet another progenitor of this relatively uncommon type of
core-collapse SNe. Very early 
light curves and spectra were interpreted by \citet{arcavi11}, using an
analytical representation of the post-shock-breakout cooling, to be
indicative of a compact progenitor, possibly a Wolf--Rayet (WR)
star. However, through 
inspection of deep pre-explosion images from the {\em Hubble Space
  Telescope} ({\em HST}), \citet{maund11} and \citet{vandyk11}
independently found an object at the SN location that was compatible
with an extended yellow supergiant star (YSG), and not with a WR
star. The question arose as to whether this object was the actual
progenitor, a companion in a binary system, or an unrelated object in
the line of sight. While \citet{vandyk11} speculated that the actual
progenitor was an unseen hot compact star, \citet{maund11} suggested
the possibility of having a YSG progenitor in a close binary system,
similar to the case of SN~1993J. By increasing the mass-loss rates in
single stellar evolution models, \citet{georgy12} was able to produce 
YSG progenitors of core-collapse SNe, although without providing a
physical motivation for the modified mass loss. Soon after,
\citet{soderberg12} suggested the progenitor must have been a compact
star based on early radio observations of the SN. The latter was
subsequently supported by \citet{krauss12} and
\citet{bietenholz12}. However, \citet{maeda12} and 
\citet{horesh13} later questioned the ability of radio data to
indicate the actual progenitor size, thus arguing against the
necessity of a compact progenitor.  

In the meantime, \citet{bersten12} performed a more detailed, numerical
hydrodynamical modeling of the SN light curve and expansion velocity,
and showed that the progenitor could not have been a massive WR star and,
very importantly, that it had a radius of $R \gtrsim 200$ $R_\odot$,
compatible with a YSG star. Bersten et~al.\ \citep[and later][in greater
  detail]{benvenuto13} presented evolutionary models of close binary
systems that naturally provided a complete picture of the
progenitor of SN~2011dh. 
Their calculations were able to simultaneously explain the explosion
of a YSG star with a final mass 
compatible with the result of the hydrodynamical modeling, with a
small amount of hydrogen in the envelope, suitable for producing a
SN~IIb, and with a companion star that would not be detected in the
pre-explosion images. 

The debate was finally settled by \citet{vandyk13} with new {\em HST}
optical imaging of the SN site obtained 21 months after explosion
\citep[this was confirmed by][using ground-based images]{ergon14a}. The
new images showed the pre-explosion object had disappeared, thus proving
that the YSG star had exploded. At the moment of those observations
the SN ejecta were too bright to determine the presence of the
companion star proposed by \citet{bersten12} and
\citet{benvenuto13}. It is only recently that it is possible to find
that last piece of the puzzle that can confirm the binary nature of
the progenitor. Observations should be carried out
in the UV range to improve the chances of detecting the hot companion.

This Letter presents deep near-UV imaging of the site of SN~2011dh
obtained with {\em HST} when the SN had an age of about 1161
days. The observations were used to search for the companion star of
the YSG progenitor, and to characterize the properties of the
progenitor system. 

\section{OBSERVATIONS AND PHOTOMETRY}
\label{sec:obs}

\begin{figure*}[htpb] 
\epsscale{0.99}
\plottwo{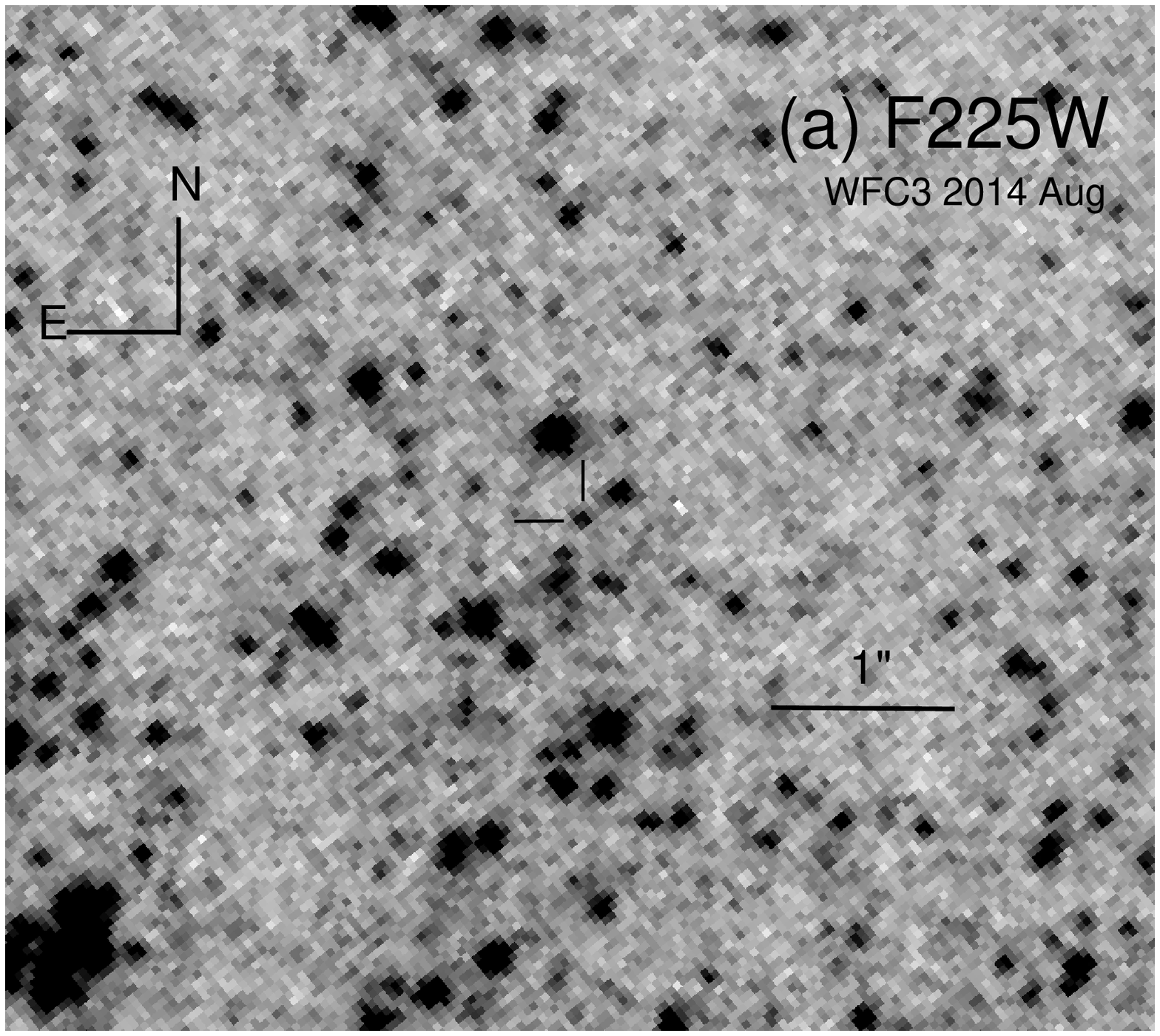}{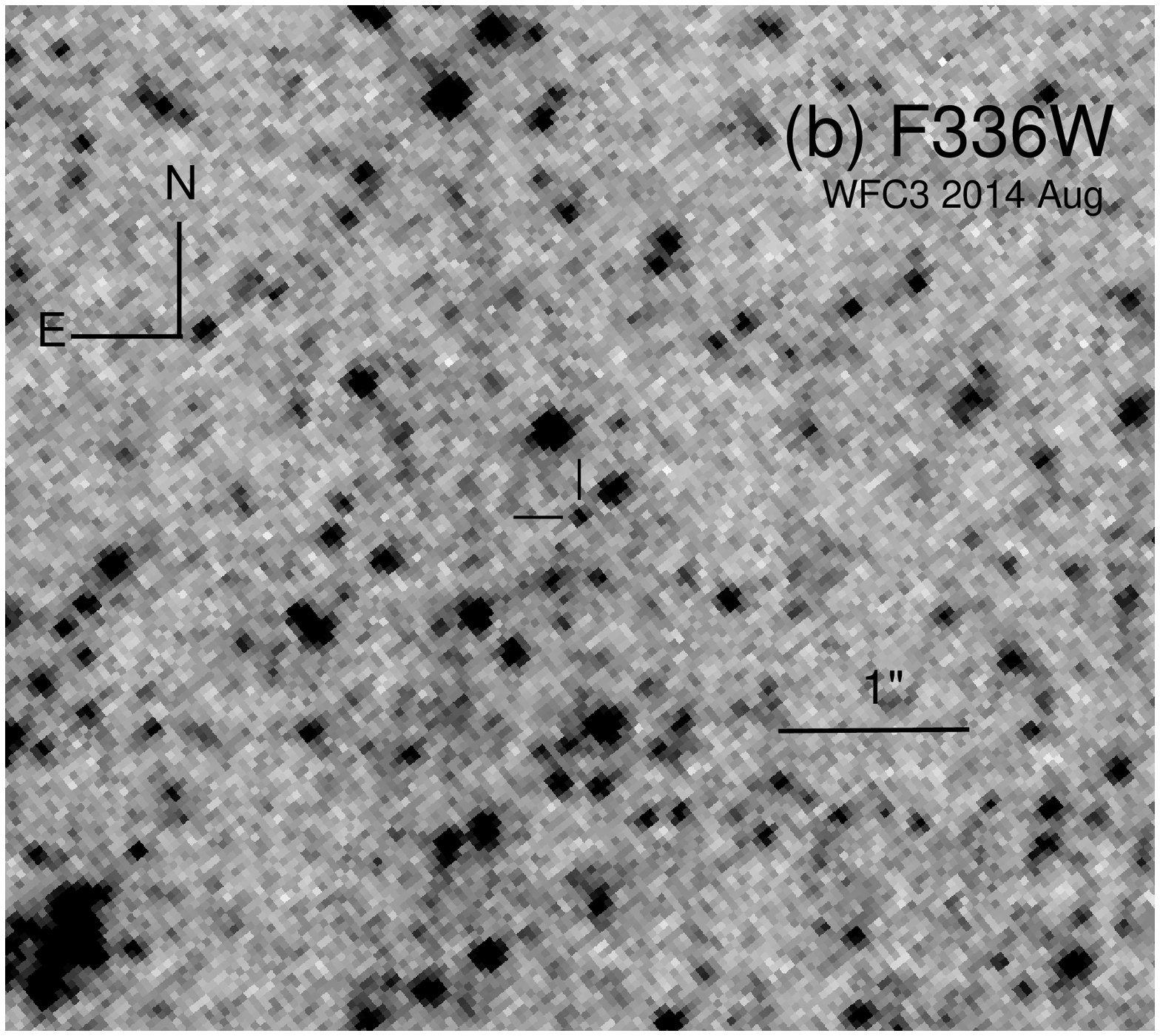}\\
\plottwo{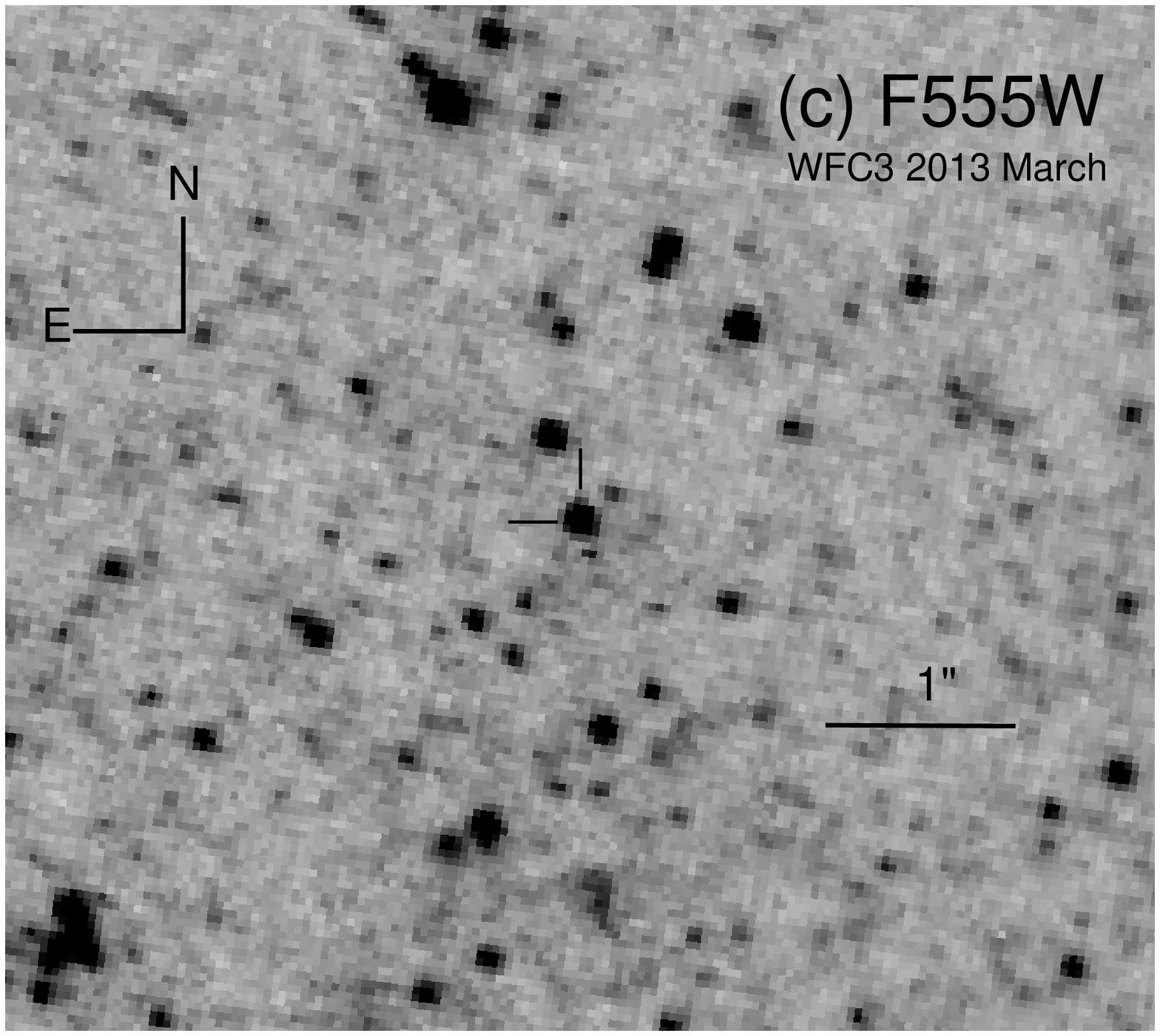}{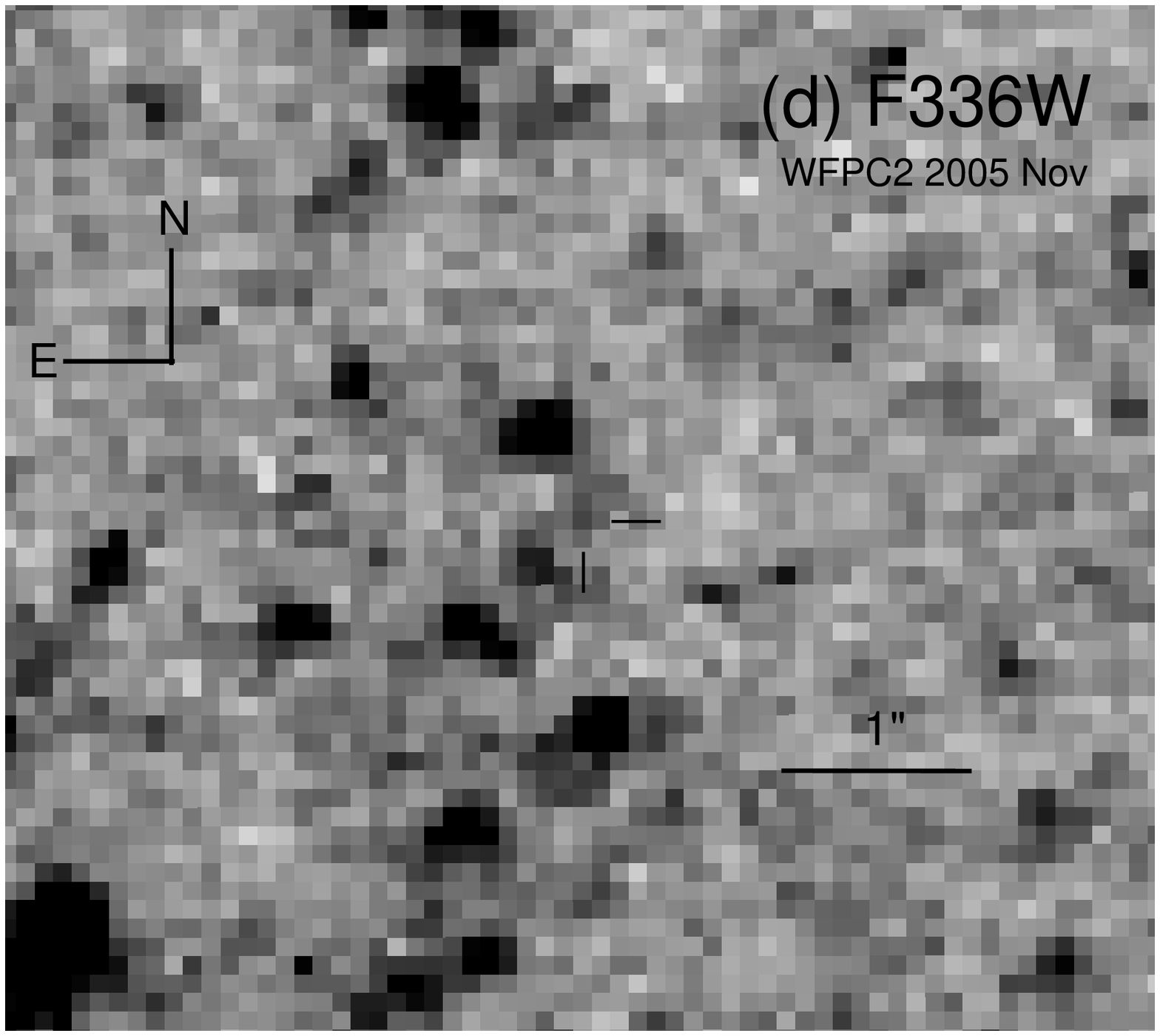}
\caption{(a) and (b) Portions of the deep F225W- and F336W-band images
  obtained with the 
  {\em HST} and WFC3/UVIS on 2014 August 7.2 UT. (c) 2013 March image
  obtained with WFC3/UVIS and filter F555W showing the fading SN
  ejecta \citep{vandyk13}. (d) Pre-explosion image obtained with the
  WFPC2 on 2005 November showing the progenitor detection
  \citep{maund11,vandyk11}. Images in all four panels are centered at
  the SN location and scale and orientation are indicated. A 
  point source is detected in both near-UV images from 2014 August, as
  indicated with tick marks. The position of the source is coincident
  with that of the SN in the 2013 March image to a projected distance
  of $\lesssim$$0.15$ pc. 
  \label{fig:img}}
\end{figure*}

\noindent The site of SN~2011dh was re-observed on 2014 August 7.2
UT\footnote{Dates are given in UT time throughout the paper.}
with {\em HST} and the Wide Field Camera 3 (WFC3) using the UVIS
channel and filters F225W and F336W. The observations were obtained
through Cycle 21 program GO-13426 (PI: J.~Maund) with which our program
GO-13433 (PI: G.~Folatelli) was merged once it was realized that both
programs shared the same target, similar science, and similar
instrumental setup. The exposure times were 3772 s in F225W, and 1784 s in
F336W. In Figure~\ref{fig:img} we show the images around the location 
of the SN. For comparison, we show a previous image obtained with
WFC3/UVIS and filter F555W on 2013 March 2 through program GO-13029
(PI: A.~V. Filippenko). This 2013 image clearly shows the fading SN,
which was also detected in a F814W-band image obtained simultaneously
\citep{vandyk13}. Figure~\ref{fig:img} also shows the pre-explosion
F336W-image obtained with the Wide Field Planetary Camera 2 (WFPC2) on
2005 November 13 (GO-10501; PI: Chandar), where the progenitor was
detected with low signal-to-noise ratio \citep{vandyk11,maund11}. 

A point source is identified at the location of the SN in both 2014
August images \citep{folatelli14}. We performed image registration of
our 2014 F336W image relative to the F555W image from 2013, using
  35 common stars. The formal rms uncertainty in the geometric
  transformation was $0.08$ pixel in each direction. The
  transformed location of the SN was offset from the blue object by
  only $0.05$ pixel in each axis. Adding both sources of
  uncertainty in quadrature, the new object is coincident with the SN
  to $0.1$ pixel, or 4 mas.
At an assumed distance to M51 of $7.8$ Mpc \citep[][from an average of
  several measurements]{ergon14a}, the new object is within $\approx$$0.15$ pc
from the SN position. We performed point-spread function (PSF)
photometry on the {\em HST} images using the Dolphot v2.0 package
\citep{dolphin00}. The resulting VEGAMAG magnitudes for the
detected object are $m_{\mathrm{F225W}}=24.57 \pm 0.11$ mag, and
$m_{\mathrm{F336W}}=24.94 \pm 0.11$ mag. 

\section{ANALYSIS}
\label{sec:ana}

\subsection{Nature of the Detected Object} 
\label{sec:ana1}

\noindent The object detected in our new {\em HST} images
coincides with the SN location, as shown in Section~\ref{sec:obs}. We
thus tested whether the detected flux could be explained by the fading
SN alone, or if it required an additional source. We estimated the
contribution from the SN ejecta at an age of 1161 days based on
previous observations of SN~2011dh, and assuming 
the light-curve decline followed that of SN~1987A at a similar
age. SN~1987A provides the richest available data set of late-time UV
observations. Even if its early-time evolution was different from that
of SN~2011dh, at the phase considered here, it is fair to assume the
light curves of both SNe are regulated by radioactive decay processes
and thus show a similar shape. We thus assumed there is no strong
interaction of the ejecta with the circumstellar medium (CSM). In
  the following we conservatively assumed only Galactic extinction of
  $A_V =0.10$ mag for SN~2011dh \citep{schlafly11}, and a total
  extinction of $A_V = 0.53 \pm 0.06$ mag for SN~1987A \citep[see][]{pun95}. A
  standard reddening law of \citet{cardelli89} was adopted, with
  $R_V=3.1$.

\begin{figure*}[htpb] 
\epsscale{0.99}
\plotone{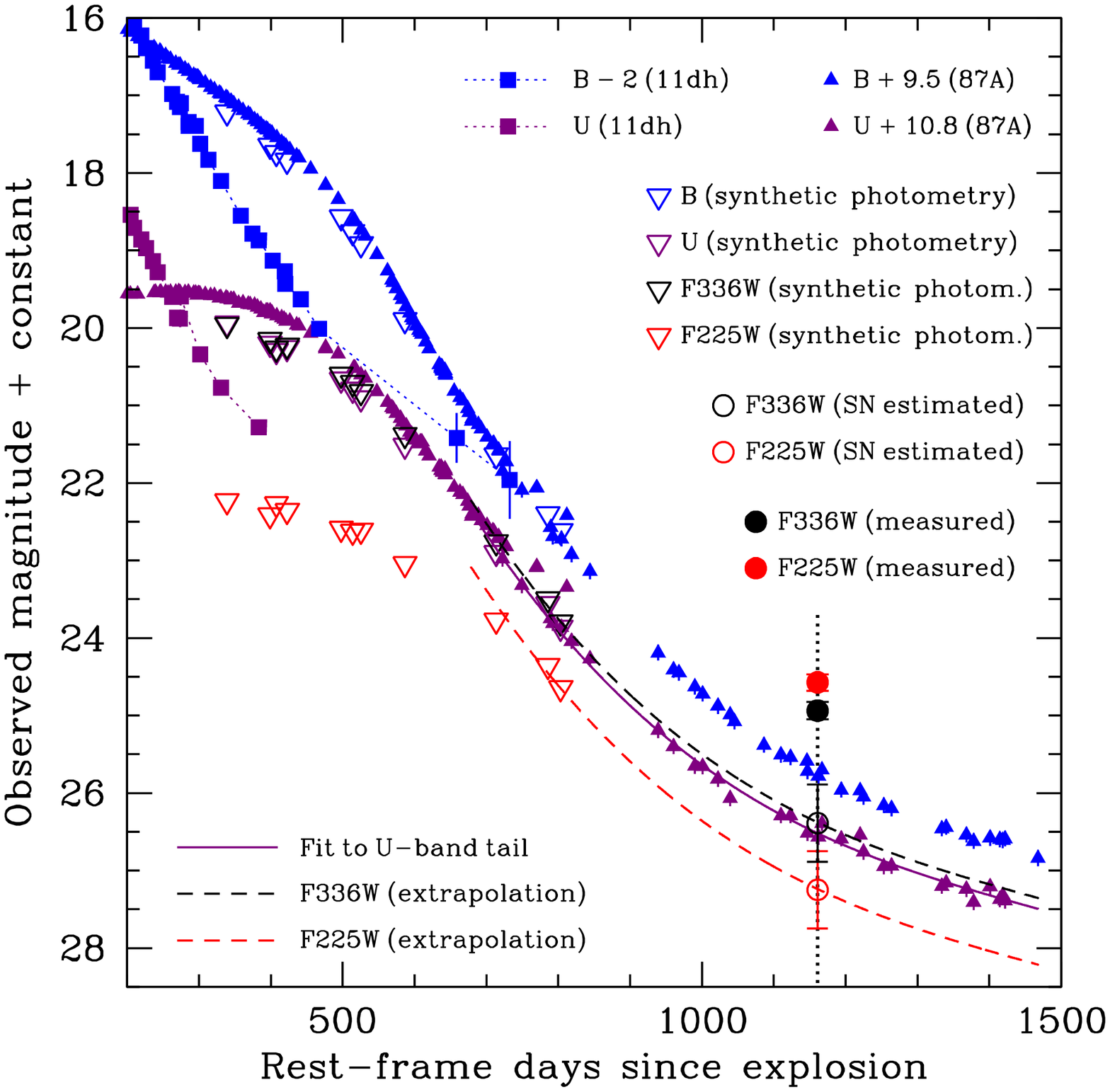}
\caption{Estimated brightness of the fading SN. Squares indicate
  the $U$- and $B$-band light curves of SN~2011dh from
  \citet{ergon14b}. Filled triangles show the light curves of SN~1987A
  \citep{hamuy90,walker91} shifted to match the evolution of
    SN~2011dh (see text). Open triangles show the synthetic 
  photometry in the {\em HST} bands computed from spectra of SN~1987A
  \citep{pun95} and shifted to the scale of SN~2011dh. The solid
    curve is a fit to the late-time decline of the $U$-band 
  light curve of SN~1987A. Dashed curves are the same fit, shifted
  vertically to match the synthetic light curves in F225W and 
  F336W bands. The vertical dotted line indicates the epoch
  of our {\em HST} observations where the SN brightness was
    estimated from the shifted decline fits.
  \label{fig:lcsextr}}
\end{figure*}

The procedure for estimating the UV flux of the SN is depicted in
Figure~\ref{fig:lcsextr}. We  
matched the $B$-band light curve of SN~1987A from \citet{hamuy90} and
\citet{walker91} to the latest photometry of SN~2011dh at $\approx$700
days from \citet{ergon14b}. From this we determined an offset of
$11.5$ mag between both SNe. Assuming the difference in
extinction given above, the offset applied in the $U$ band was of $10.8$
mag. We then calculated synthetic photometry
in the $U$, $B$, F225W, and F336W bands using the spectra of SN~1987A
published by \citet{pun95}, after correcting for the difference
  in reddening. The synthetic photometry was roughly consistent
with the observed $U$- and $B$-band photometry, which provides
confidence for its use to derive F225W and F336W magnitudes. The
spectra of SN~1987A extended until about 800 days after
explosion. There is a final spectrum obtained at about 1050 days, but
it is too noisy in the near-UV range to be used here. 

We extrapolated the synthetic photometry in F225W and
F336W, to the epoch of our {\em HST} observations. For this purpose,
we used a cubic polynomial fit to the $U$-band light curve of SN~1987A
between 700 and 1500 days, and assumed the F225W and F336W bands
followed the same decline rate. With this, the extrapolated magnitudes
of the SN ejecta in the {\em HST} bands were
$m_{\mathrm{F225W}}(1161\,\mathrm{days})=27.3 \pm 0.5$ mag, and 
$m_{\mathrm{F336W}}(1161\,\mathrm{days})=26.4 \pm 0.5$ mag. The uncertainties were
estimated based on the extinction uncertainty, and on the
variation in ($U$--near-UV) colors among the latest spectra of
SN~1987A. This analysis suggests that in the absence of strong CSM
interaction, the SN was $\approx$$2.7$ mag and $\approx$$1.5$
  mag fainter than the observed object in F225W and F336W,
respectively. 

The emission from the expanding SN ejecta may be revitalized if it
encounters substantial circumstellar material and a shock is
produced. This would cause a flattening of the light curve and it
would introduce signatures in the spectrum. No evidence of strong
interaction was seen in the optical light curves or spectra of
SN~2011dh at 700 days after explosion
\citep{shivvers13,ergon14b,jerkstrand14}, in X-rays at
$\approx$500 days \citep{maeda14}, or in radio at $\approx$ 1000 days
  (A.\ Kamble, private communication).  Contrary to SN~1993J which
showed signs of interaction in this wavelength range 
as soon as 100 days after explosion
\citep[e.g.,][]{filippenko94,kohmura94}, SN~2011dh appears 
to be comparatively free of CSM \citep{maeda14}. Even if CSM
  interaction strengthened after $\approx$1000 days, it would be
  difficult to explain the observed blue UV color. Spectra of SN~1993J
  at similar ages \citep{fransson05} show no strong lines in the range
  of the F225W filter that could be responsible for the observed color.

An alternative explanation to the observed flux may be an unresolved
echo of the SN light reflected in our direction. Any such echo should
arise from within $\sim$1 pc from the SN. At that distance, we only
expect to find CSM created from the pre-SN mass loss, which can be
assumed to have too low a density to produce a detectable echo. In
addition, light from the SN should be backward-scattered, which is
relatively inefficient in the near-UV \citep{sugerman03}.

It is in principle possible that the observed emission is due to an
unrelated source in the line of sight. However, we consider this
unlikely because such an object, presumably an OB
type star, should be projected at $\lesssim$$0.15$ pc from the SN,
which leaves a small volume across the disk of the face-on galaxy M51.

\subsection{Putative Companion Star Properties} 
\label{sec:ana2}

\begin{figure*}[htpb] 
\epsscale{0.99}
\plotone{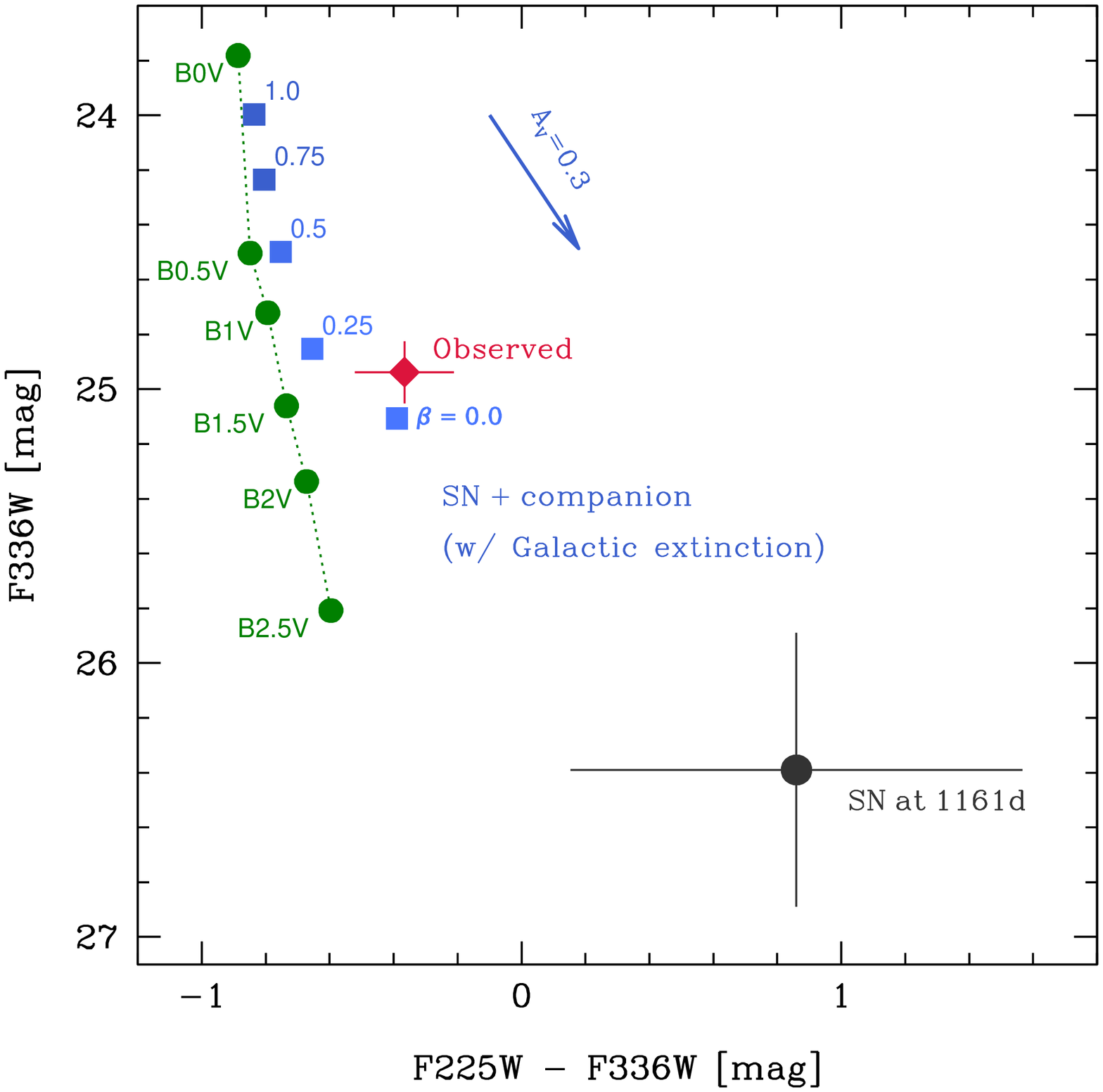}
\caption{F336W versus $(\mathrm{F225W}-\mathrm{F336W})$
  color--magnitude diagram. The observed object is shown with a red
  diamond, without correcting for extinction. The black dot shows the
  expected location of the SN, as estimated in
  Figure~\ref{fig:lcsextr}. Blue squares indicate the expected
  observations from the SN plus a binary companion
    (SN\,+\,companion) according to the 
  models of \citet{benvenuto13}, for varying mass-accretion
  efficiency ($\beta$; labeled next to each point). A distance of
  $7.8$ Mpc was adopted, and the SN\,+\,companion model photometry was
  reddened by Galactic extinction of $A_V=0.10$ mag to reproduce the
  observations. Green dots show the location of main sequence stars
  for the same distance and reddening. The arrow indicates the
  reddening vector for an additional extinction of $A_V=0.3$ mag that may
  be produced by dust in M51 (see the text for details).
  \label{fig:colmag}}
\end{figure*}

\noindent From the analysis in the previous section it is likely that
the detected object is the companion of the YSG progenitor of
SN~2011dh. If this is the case, its observed near-UV color and 
magnitude can be used to characterize the companion star, as shown in
Figure~\ref{fig:colmag}. Adopting a Galactic extinction only and an 
average distance of $7.8 \pm 0.9$ Mpc, the 
absolute magnitudes of the object are $M_{\mathrm{F225W}}=-5.11 \pm
0.29$ mag, and $M_{\mathrm{F336W}}=-4.66 \pm 0.29$ mag. Extinction by
dust in M51 is uncertain. \citet{arcavi11} and \citet{ritchey12} used
the strength of Na\,I~D lines in the SN spectra to suggest that
extinction was low. However, from color analysis of the stellar
population in the vicinity of the SN, \citet{murphy11} found extinction of 
$A_V \approx 0.3$ mag. Assuming the latter value, the corrected
  absolute magnitudes would be $M_{\mathrm{F225W}}=-5.88$ mag and
$M_{\mathrm{F336W}}=-5.15$ mag. The allowed range of absolute
magnitudes is roughly compatible with main-sequence, B0--B2 type
stars. 

There is the possibility that additional dust is formed in the SN
ejecta, which would introduce extra extinction. \citet{nozawa10}
predict that significant amounts of dust can form in the ejecta of
Type~IIb SNe approximately between 300 and 700 days after the explosion,
but that the average grain size is smaller than in Type II~P SNe. In
addition, UV emission from a putative hot star present within the
ejecta can destroy the small dust grains. Dust formation has been
suggested for SN~2011dh as early as 150 days 
\citep{sahu13,ergon14b,jerkstrand14}, however it is difficult to
quantify the amount of extinction in the near-UV regime. If
present, additional dust would imply an intrinsically more luminous
and bluer object than what was derived above. 

Figure~\ref{fig:colmag} also shows the {\em HST} photometry compared with
the predictions from the 
close-binary evolutionary models of \citet{benvenuto13}. The suggested
progenitor system had zero-age masses of 16 $M_\odot$ (the exploding
star) and 10 $M_\odot$ (the accreting companion), and an initial orbital
period of 125 days. Depending on the value of the mass-accretion
efficiency parameter, $\beta$, the model predicted different final
masses, luminosities, and temperatures for the companion. The companion star
spectra were adopted from atmosphere models of \citet{kurucz93}
  with an effective temperature and surface gravity as given by the
  binary models and were scaled to the distance of M51. To this we added an
  SN contribution using the spectrum 
  of SN~1987A at $\approx$800 days, dereddened, and scaled to the
  expected at 1161 days as estimated in Section~\ref{sec:ana1}
  (Figure~\ref{fig:lcsextr}). The resulting ``SN\,+\,companion''
  spectra were reddened by 
  Galactic extinction and synthetic photometry was computed in F225W
  and F336W bands. As shown in Figure~\ref{fig:colmag}, the observed
  object is remarkably compatible with the SN\,+\,companion photometry
  provided the accretion efficiency is low.
For comparison, the diagram shows the estimated position of the
SN ejecta. Assuming additional extinction in the host galaxy of
$A_V \lesssim 0.3$ mag, the proposed companion star is compatible
with the observations for $\beta \lesssim 0.5$. This implies a mass range of
$10 \lesssim M \lesssim 16$ $M_\odot$, for the secondary star at the
time of the SN explosion. Note, however, that we expect a certain degree of
degeneracy in the models between $\beta$ and the initial mass of the
companion, which would require more extensive observations and
modeling to disentangle the degeneracy. If $\beta$ is small, one
  can expect part of the mass lost by the primary to be in the
  surroundings and to eventually be shocked by the ejecta. In the
  models of \citet{benvenuto13}, however, the mass-transfer rate is
  high ($\sim$$10^{-3}$ $M_\odot$ yr$^{-1}$) only at $\sim$$10^5$ yr
  before the explosion, so we can assume the ejecta (expanding at
  $\sim$$10^4$ km s$^{-1}$) will reach this material (if expelled at
  $\sim$10 km s$^{-1}$) not earlier than $\sim$$10^2$ yr.   

The predicted
  SN\,+\,companion photometry for the optical {\em HST} observations
  obtained on 2014 August 10 (GO-13426) are
$25.8 \lesssim m_{\mathrm{F435W}} \lesssim 26.6$ mag, 
$26.0 \lesssim m_{\mathrm{F555W}} \lesssim 26.7$ mag, and 
$26.0 \lesssim m_{\mathrm{F814W}} \lesssim 26.4$ mag.
If there is additional dust formed in the ejecta, the object would be
intrinsically brighter, which would imply a larger companion
mass, presumably due to a larger accretion efficiency. 

\section{CONCLUSIONS}
\label{sec:concl}

\noindent We have shown the detection in deep near-UV images obtained
with {\em HST} of a blue point source at the location of the Type~IIb
SN~2011dh. The source's photometry is compatible with it being
the progenitor companion predicted by \citet{bersten12} and 
\citet{benvenuto13}. We
consider it unlikely that the observed flux is due to the SN ejecta
itself, to a light echo, or to an unrelated object in the line of
sight. SN~2011dh would thus be the second core-collapse SN, after 
  SN~1993J \citep{maund04,fox14}, to show strong evidence of a
  binary companion to the progenitor. If confirmed, the companion of
  SN~2011dh would already be the dominant source in the optical--UV
  regime, thus providing a unique opportunity for analyzing its properties.

These observations may provide important clues about
the evolutionary paths of massive stars, and the role
of binarity in the envelope removal among H-poor SNe.
The fact that the two best-studied Type~IIb SNe most probably had
binary progenitors is particularly suggestive of a dominant binary
channel for this subtype of SNe, especially considering the difficulty
of single stellar evolution models to produce progenitors that lose
most but not all of their H-rich envelopes. 

Our analysis allowed us to provide a range of valid mass-accretion
efficiency $\beta \lesssim 0.5$ for the specific models presented by
\citet{benvenuto13}. In general, this quantity is included in the
models as a free, unconstrained parameter. However, a conclusive derivation
of the binary model properties would require a more detailed study and
further observational data covering a wider wavelength range. At the
same time, such observations may allow us to better determine the
degree of interaction between the ejecta and the CSM, the possible
contamination from a light echo, and the amount of extinction from
pre-existing and newly formed dust. 

\acknowledgments 
We thank Armin Rest for his light-echo calculations and
discussion. This research is supported by the World Premier
International Research Center Initiative (WPI Initiative), MEXT, Japan, and
by Grant-in-Aid for Scientific Research (23224004, 23540262,
23740141, 23740175, 26400222, 26400223, 26800100).
M.H. and H.K.\ acknowledge support from the
Millennium Institute of Astrophysics (MAS; 
Programa Iniciativa Cient\'ifica Milenio del Ministerio de Econom\'ia,
Fomento y Turismo de Chile, grant IC120009). H.K.\ is supported by FONDECYT 
(grant 3140563). O.G.B.\ is member of the Carrera
del Investigador Cient\'ifico of CIC, Argentina.

\end{document}